\documentstyle[12pt,aasms4]{article}

\def\vA{v_{\scriptscriptstyle A}}
\def\Emin{E_{\scriptstyle min}}
\def\Emax{E_{\scriptstyle max}}
\def\Eqn#1{Eqn (\ref{eq:#1})}

\begin{document}

\title{Turbulent coronal heating and the distribution of nanoflares} 
\author{Pablo Dmitruk\altaffilmark{1} and Daniel O. G\'omez\altaffilmark{2,3}}
\affil{Departamento de F\'\i sica, Facultad de Ciencias Exactas y 
Naturales, \\ Universidad de Buenos Aires,\\ Ciudad Universitaria, 
Pabell\'on I, 1428 Buenos Aires, Argentina}

\altaffiltext{1}{Fellow of Universidad de Buenos Aires}
\altaffiltext{2}{Also at the Instituto de Astronom\'\i a y F\'\i sica
del Espacio, Buenos Aires, Argentina}
\altaffiltext{3}{Member of the Carrera del Investigador, CONICET, Argentina}

\authoremail{dmitruk@df.uba.ar}

\begin{abstract}
We perform direct numerical simulations of an externally driven two-dimensional
magnetohydrodynamic system over extended periods of time to simulate the 
dynamics of a transverse section of a solar coronal loop. A stationary and 
large-scale magnetic forcing was imposed, to model the photospheric motions at 
the magnetic loop footpoints.  
A turbulent stationary regime is reached, which corresponds
to energy dissipation rates 
consistent with the heating requirements of coronal loops.

The temporal behavior of quantities such as the energy dissipation rate show 
clear indications of intermittency, 
which are exclusively due to the strong nonlinearity of 
the system.  
We tentatively associate these impulsive events of magnetic energy dissipation 
(from $5 \times 10^{24}$ erg to $10^{26}$ erg)
to the so-called nanoflares. A statistical analysis of these events yields
a power law distribution as a function of their energies with a negative
slope of $1.5$, which is consistent with those 
obtained for flare energy distributions reported from X-ray observations.
\keywords { Sun: flares --- MHD --- turbulence} 
\end{abstract}

\section{Introduction}

Coronal loops in active regions are likely to be heated by Joule dissipation 
of highly structured electric currents. The formation of small scales in the 
spatial distribution of electric currents is necessary to enhance magnetic 
energy dissipation and therefore provide sufficient heating to the plasma 
confined in these loops. Various scenarios of how these fine scale structures 
might be generated have been proposed, such as the spontaneous formation of 
tangential discontinuities (\cite{Par72}, \cite{Par83}), the development of an 
energy cascade driven by random footpoint motions on a 
force-free configuration (\cite{van86}), or the direct energy cascade 
associated to a fully turbulent magnetohydrodynamic (MHD)
regime (\cite{Hey92}, \cite{Gom92}). These 
rather different models share in common the dominant role of nonlinearities in 
generating fine spatial structure.

In this paper we assume that the dynamics of a coronal loop
driven by footpoint motions is described by 
the MHD equations.
Since the kinetic ($R$) and magnetic ($S$) Reynolds 
numbers in coronal active regions are extremely large ($R \sim S \sim 
10\sp{10-12}$), we expect footpoint motions to drive the loop into a 
strongly turbulent MHD regime.

Footpoint motions whose lengthscales are much smaller than the 
loop length cause the coronal plasma to
move in planes 
perpendicular to the axial magnetic field, generating a small transverse 
magnetic field component.
In \S 2 we model this coupling to 
simulate the driving action of footpoint motions on a generic transverse 
section of the loop. The numerical technique used for the integration of the 
two-dimensional
MHD equations is described in \S 3. In \S 4 we report the energy dissipation 
rate that we obtain, and a statistical analysis of dissipation events
is presented 
in \S 5. Finally, the relevant results of this paper are summarized in \S 6.

\section{Forced two-dimensional magnetohydrodynamics}

The dynamics of a coronal loop with 
a uniform magnetic field ${\bf B} = B_0 {\bf z}$, 
length $L$ and 
transverse section $(2\pi l)\times (2\pi l)$, can be modeled by the 
RMHD equations (\cite{Str76}):
\begin{equation}
     \partial_t a  = \vA \partial_z \psi + [ \psi , a ] + 
     \eta \nabla \sp2 a 
\label{eq:rmhda}
\end{equation}
\begin{equation}
     \partial_t w = \vA \partial_z j + [ \psi , w ] - [ a , j ] +
     \nu \nabla \sp2 w
\label{eq:rmhdw}
\end{equation}
\noindent
where $\vA = B_0/\sqrt{4\pi\rho}$ is the Alfven speed, $\nu$ is the kinematic 
viscosity, $\eta $ is the plasma resistivity, $\psi$ is the stream 
function, $a$ is the vector potential, $w=-\nabla\sp2\psi\ $ is the fluid 
vorticity, $j=-\nabla\sp2 a$ is the electric current density 
and $[u,v]={\bf z} \cdot {\bf \nabla}u \times {\bf \nabla}v$.  
For given photospheric motions 
applied at the footpoints (plates $z=0$ and $z=L$) horizontal velocity and 
magnetic field components develop in the interior of the loop, given by ${\bf 
v} = \nabla\times ({\bf z}\psi)$ and ${\bf b} = \nabla\times ({\bf z}a)$. 

The RMHD equations can be regarded as describing a set of two-dimensional MHD 
systems stacked along the loop axis and interacting among themselves through 
the $\vA\partial_z\ $ terms. For simplicity, hereafter we study the evolution 
of a generic two-dimensional slice of a loop to which end we model the 
$\vA\partial_z\ $ terms as external forces (see \cite{Ein96} for a similar 
approach). We assume the vector potential to be independent of $z$ and the 
stream function to interpolate linearly between $\psi (z=0) = 0$ and $\psi 
(z=L) = \Psi $, where $\Psi (x,y,t)$ is 
the stream function for the photospheric velocity field. 
These assumptions yield $\vA \partial_z\psi = \vA\Psi /L$ (in 
\Eqn{rmhda}) and $\vA \partial_z j = 0$ (in \Eqn{rmhdw}) and 
correspond to an idealized scenario where
the magnetic stress distributes uniformly throughout  
the loop. The 2D equations for a generic transverse slice 
of a loop become,
\begin{equation}
     \partial_t a = [ \psi , a ] + \eta \nabla \sp2 a + f
\label{eq:dta}
\end{equation}
\begin{equation}
     \partial_t w = [ \psi , w ] - [ a , j ] + \nu \nabla \sp2 w
\label{eq:dtw}
\end{equation}
\noindent 
where $f = (\vA/L)\Psi$.

\section{Numerical procedures}

We performed numerical simulations of Eqs (\ref{eq:dta})-(\ref{eq:dtw}) on 
a square box of sides 
$2\pi$, with periodic boundary conditions.  The magnetic vector potential and 
the stream function are expanded in Fourier series.
To be able to perform long-time integrations, we worked with a moderate 
resolution version of the code ($96 \times 96$ grid points).
The code is of the pseudo-spectral 
type, with $2/3$ de-aliasing (\cite{Can88}). The temporal integration scheme 
is a fifth order 
predictor-corrector, to achieve almost exact energy balance  
over our extended time simulations.

We turn Eqs (\ref{eq:dta})-(\ref{eq:dtw}) into a dimensionless version, 
choosing $l$ and $L$ as the units for transverse and longitudinal distances, 
 and $v_0 = \sqrt{f_0}$ as the unit for velocities
($f_0 = \vA u_p (l/L)$, $u_p$: typical photospheric 
velocity), since the field intensities are 
determined by the forcing strength. The dimensionless dissipation coefficients 
are $\nu_0=\nu /(l~\sqrt{f_0})$ and $\eta_0=\eta /(l~\sqrt{f_0})$. The forcing 
is constant in time and non zero only in a narrow band in $k$-space
corresponding to $3 \le k~l \le 4$.
In spite of the narrow forcing and even though velocity and magnetic fields 
are initially zero, non-linear terms quickly populate all the modes across 
the spectrum and a stationary turbulent state is reached.

\section{Energy dissipation rate}

To restore the dimensions to our numerical results, we used typical values 
for the solar corona: $L \sim 5 \times 10\sp{9}{\rm cm}$,
$l \sim 10\sp{8}{\rm cm}$, $v_p \sim 
10\sp{5}{\rm cm}/{\rm s}$, $B_0 \sim 100~{\rm G}$, 
$n \sim 5 \times 10^9\ {\rm cm}^{-3}$ and
$\nu_0=\eta_0=3 \times 10\sp{-2}$.

Fig. 1 shows magnetic and kinetic energy vs time. The behavior of both 
energies is highly intermittent despite the fact that the forcing is constant 
and coherent. This kind of behavior is usually called internal intermittency, 
to emphasize the fact that the rapid fluctuations are not induced by an 
external random forcing. Also, note that magnetic energy is about one order of 
magnitude larger than kinetic energy. 
\placefigure{f1}
The energy dissipation rate is also a strongly intermittent quantity as 
shown in Fig. 2. For turbulent systems at large Reynolds numbers, the 
dissipation rate in the stationary regime is expected to be independent 
of the Reynolds number R (\cite{Kol41}). For the rather moderate 
Reynolds number simulations reported here, a weak (monotonically 
increasing) dependence of the dissipation rate with R still remains (see 
also \cite{Ein96}).
\placefigure{f2}
The total dissipation rate for the 
typical numbers  
listed above is $\epsilon\simeq 1.6 \times 10^{24}\ {\rm erg.s}^{-1}$. 
We can transform the heating rate into an energy 
influx from the photosphere, by simply dividing by twice the transverse 
area (because we have two boundaries), i.e. ${\cal F} = \epsilon /(2(2\pi 
l)^2)$. In \Eqn{flux} we show the quantitative value of 
the energy flux as well as the 
explicit dependence with the relevant parameters of the problem,
\begin{equation}
 {\cal F} = 2 \times 10^6 {{\rm erg}\over{{\rm cm}^2~{\rm s}}}
 \left({n\over{5 \times 10^9~{\rm cm}^{-3}}} \right) ^{1\over4}
 \left( {B_0\over{100~{\rm G}}} \right) ^{3\over2}
 \left( {u_{ph}\over{10^5~{\rm cm}~{\rm s}^{-1}}} \right) ^{3\over2}
 {{({l_{ph}\over{10^8~{\rm cm}}})^{1\over2}}\over{
 ({L\over{5 \times 10^9~{\rm cm}}})^{1\over2}}}
\label{eq:flux}
\end{equation}
\noindent
This energy flux compares quite favorably with the heating requirements 
for active regions, which span the range ${\cal F} = 3 \times 10^5\ -\ 
10^7~{\rm erg}~{\rm cm}^{-2}~{\rm s}^{-1}$ (\cite{Wit77}).

\section{Distribution of nanoflares}

In this section, we associate the peaks of energy dissipation displayed 
in Fig. 2 to the so-called nanoflares (\cite{Par88}). We estimate 
the occurrence rate for these nano-events, i.e. the number of events 
per unit energy and time $P(E)=dN/dE$ so that,
\begin{equation}
{\cal R} = \int^{\Emax}_{\Emin} dE~P(E)
\label{eq:number}
\end{equation}
\noindent
is the total number of events per unit time and
\begin{equation}
\epsilon = \int^{\Emax}_{\Emin} dE~E~P(E)
\label{eq:heat}
\end{equation}
\noindent
is the total heating rate (in ${\rm erg}~{\rm s}^{-1}$) contributed by all  
events in the energy range $[\Emin\ ;\ \Emax\ ]$. A simple inspection 
to the $\epsilon (t)$ time series shown in Fig. 2 indicates that these 
events are in a highly concentrated or piled-up regime, i.e. their event 
rate ${\cal R}$ multiplied by their typical duration is much larger than 
unity. At any given time, 
many events are going on simultaneously.

This piled-up scenario is a serious drawback against performing any 
kind of statistical analysis. 
To overcome this difficulty we define an event in the 
following fashion: (1) we set a threshold heating rate $\epsilon_0$ on the 
time series displayed in Fig. 2, of the order of its time average, 
(2) events are excesses of dissipation which start when
$\epsilon (t)$ surpasses $\epsilon_0$ and finish when $\epsilon (t)$
returns below $\epsilon_0$.
Once a particular threshold is set, we perform a statistical analysis
of the events, keeping track of their peak values, durations and total 
energy content. The implicit assumption behind our working definition,  
is that the small fraction of events that emerge over the 
threshold are statistically representative of the whole set.  We do not 
make any attempt to prove this assertion, which therefore remains 
as a working hypothesis.

Among the interesting results of this statistical analysis, we 
find a significant correlation between the energy and duration 
of events like $E\simeq \tau^2$ (\cite{Nos97}), which is consistent with the 
correlation reported by \cite{Lee93} from hard X-ray observations 
and by \cite{Lu93} for an avalanche model of 
flare occurrences. Perhaps the most important result is that 
the occurrence rate as a function of energy displays a power 
law behavior
\begin{equation}
   P(E) = A\ E^{-1.5 \pm 0.2}
\label{eq:power}
\end{equation}
\noindent
in the energy range spanned from $\Emin \simeq 5 \times 10^{24}~{\rm erg}$ to
$\Emax \simeq 10^{26}~{\rm erg}$.  

We define the constant $A$ in \Eqn{power} 
so that the heating rate computed from \Eqn{heat} matches the 
total heating rate from our simulation (see \Eqn{flux}).
Figure 3 shows our occurrence 
rate, displaying the power law behavior indicated in \Eqn{power}. 
For comparison, we also plotted the occurrence rate for transient 
brightenings derived by \cite{Shi95} (slope between 1.5-1.6)
from Yohkoh soft X-ray observations 
and by \cite{Cro93} (slope 1.53) from SMM hard X-ray data.
\placefigure{f3}
Also, we obtained the distribution of events as a function of peak fluxes 
which is a power law with slope $1.7 \pm 0.3$.
This slope is consistent
with the one derived by \cite{Cro93} (1.68) from X-ray events
and somewhat flatter than those reported by \cite{Hud91} (1.8).
Note that the slopes derived from X-ray observations assume that
the luminosity in X-rays is proportional to the dissipated energy.
\cite{Por95} criticize this assumption after comparing the UV and X-ray
emission for several microflares and find that slopes derived
from X-rays become slightly steeper when corrected for
this effect.

The remarkable correspondence between the rates plotted in Fig. 3, is 
indicative of the presence of a common physical process behind the 
dissipation events ranging from $10^{24}~{\rm erg}$ or less,
to $10^{33}~{\rm erg}$ 
for the largest flares. Since in all cases the index of the power law 
remains smaller than two, \Eqn{heat} implies that the contribution 
to energy dissipation in a given energy range ($[\Emin ;\Emax ]$) 
is dominated by the most energetic events (i.e. $E\simeq \Emax$). 
According to this result, the relatively infrequent large energy events 
contribute more to the heating rate than the much more numerous 
small energy events. This assertion might only change if indications 
of a turn-up of the slope (to an index larger than two) are found 
at low energies (\cite{Hud91}). 

\section{Discussion and conclusions}

In the present paper we simulate the dynamics of a transverse section of a 
solar coronal loop through an externally driven two-dimensional MHD code.
The relevant results of this study are summarized as follows:
(1) for an external forcing which is narrowband in wavenumber, we find that 
the system becomes strongly turbulent and after about twenty photospheric 
turnover times a stationary turbulent regime is 
reached,   
(2) the energy dissipation rate obtained for typical 
footpoint velocities is consistent with the power necessary to heat active 
region loops
(${\cal F}\simeq 2 \times 10^6~{\rm erg}~{\rm cm}^{-2}~{\rm s}^{-1}$),
(3) the energy dissipation rate displays a highly 
intermittent behavior, which is a ubiquitous characteristic of turbulent 
systems, (4) we associate the elementary events that compose this intermittent 
heating rate to the so-called nanoflare events described by 
\cite{Par88}.
A statistical analysis of the events performed on a long term
numerical simulation
(about $200$ turnover times) shows a power law event rate, going like
$dn/dE \sim E\sp{-1.5}$, which is remarkably consistent with the statistics of
flare occurrence derived from observations (\cite{Hud91}, \cite{Cro93},
\cite{Lee93}, \cite{Shi95}).

\acknowledgements
We acknowledge financial support by the University of Buenos Aires 
(grant EX247) and by Fundaci\'on Antorchas.

\newpage

\figcaption[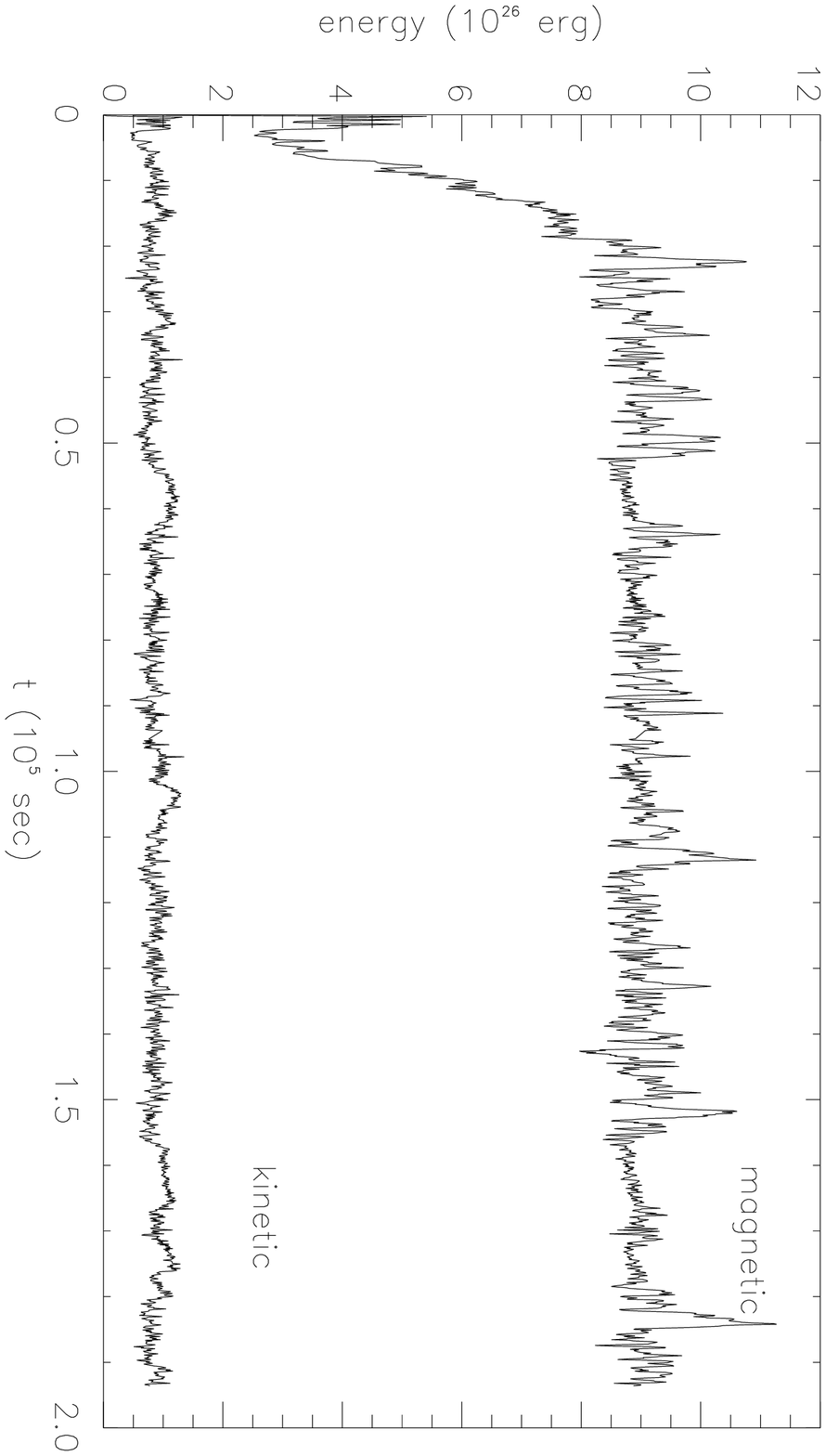]{
Magnetic and kinetic energy as functions of time for our
simulation.
\label{f1}}
\figcaption[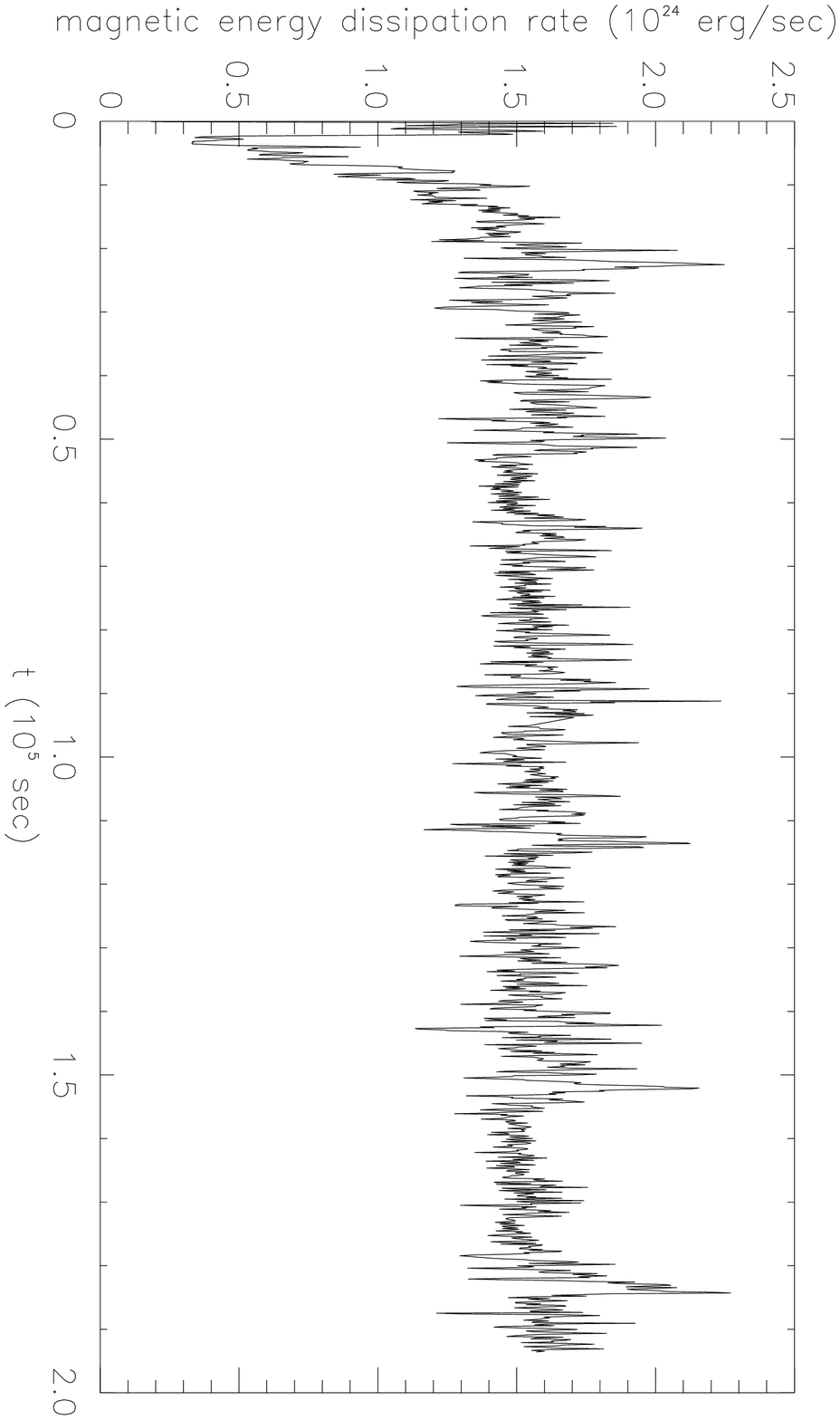]{
Magnetic energy dissipation rate as a function of time.
\label{f2}}
\figcaption[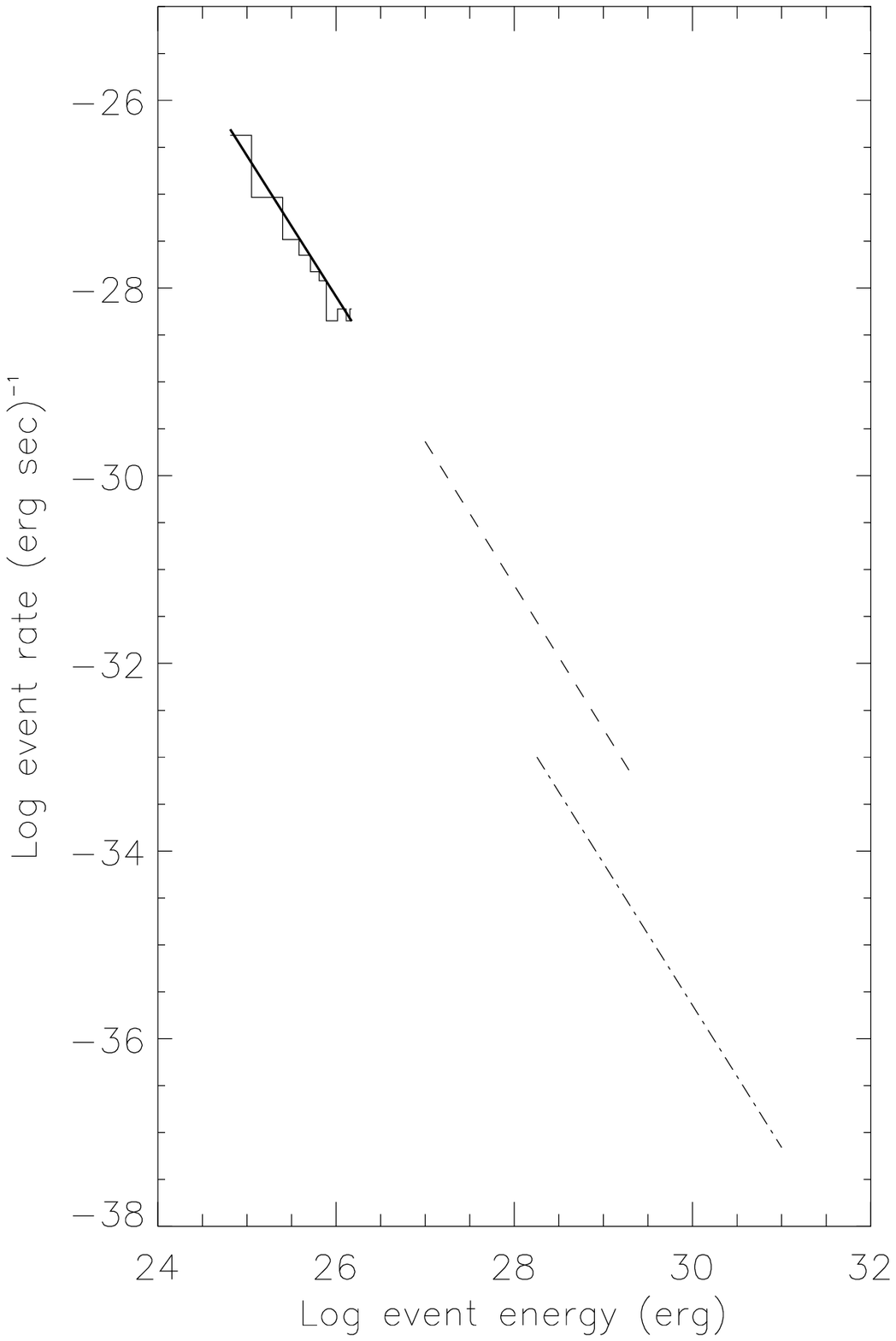]{
Number of events per unit energy and time. In the upper left corner
we plotted the histogram for our
simulation and 
the thick line shows the best power law fit (slope $1.5 \pm 0.2$).
Dashed trace correspond to the 
occurrence rates derived by Shimizu 1995, and the dot-dashed trace to the 
distributions computed by Crosby et al. 1993 
\label{f3}}

\end{document}